\documentclass[letterpaper, 10 pt, conference]{ieeeconf} 

\IEEEoverridecommandlockouts
\usepackage{cite}
\usepackage{amsmath,amssymb,amsfonts}
\usepackage[ruled, lined, longend, linesnumbered]{algorithm2e}
\usepackage{algpseudocode}
\usepackage{graphicx}
\usepackage{textcomp}
\usepackage[left=1.4cm, right=1.4cm, top=1.7cm]{geometry}
\usepackage{caption}
\usepackage{anyfontsize}
\usepackage{url}
\usepackage{dsfont}
\usepackage{subcaption}
\usepackage{soul}
\usepackage{bm}
\usepackage{xcolor}
\usepackage{multirow}
\usepackage{threeparttable}
\usepackage{BOONDOX-uprscr}
\usepackage[T1]{fontenc}
\usepackage{xspace}
\newcommand{\arxiv}{a\kern-.1667em\lower.5ex\hbox{R}\kern-.125emX\spacefactor\@}

\def\BibTeX{{\rm B\kern-.05em{\sc i\kern-.025em b}\kern-.08em
    T\kern-.1667em\lower.7ex\hbox{E}\kern-.125emX}}

\title{\LARGE \bf
	Obstacle-Aware Four-Dimensional Trajectory Design for Urban Air Mobility*}
	
\author{Prasad Devkar$^{1}$, Yashovardhan S. Chati$^{2}$, and Arunchandar Vasan$^{3}$
	\thanks{*This work was not supported by any organization except the authors' employer Tata Consultancy Services Limited.}
	\thanks{$^{1}$ Prasad Devkar is a Researcher with the Corporate Research and Innovation Division of Tata Consultancy Services Limited, India.
		{\tt\small prasad.devkar@tcs.com}}%
	\thanks{$^{2}$ Yashovardhan S. Chati is a Senior Scientist at the same division as above.
		{\tt\small ys.chati@tcs.com}}%
	\thanks{$^{3}$ Arunchandar Vasan is a Principal Scientist at the same division as above.
		{\tt\small arun.vasan@tcs.com}}%
}

\begin{document}
	
\maketitle
\thispagestyle{empty}
\pagestyle{empty}

\begin{abstract}
Urban Air Mobility (UAM) with electric Vertical TakeOff and Landing (eVTOL) vehicles can help address ground traffic congestion. The design of an eVTOL trajectory that is safe and reduces travel time is key for UAM adoption. Existing works on trajectory design either may not adequately incorporate dense obstacles in urban environments, complex eVTOL flight dynamics, or one or more flight phases. Not considering these factors can result in low-quality, or worse infeasible, trajectories. 
We develop a hybrid framework that can integrate building obstacles data, wind data, eVTOL flight dynamics, and other real-world operational constraints to estimate a four-dimensional eVTOL flight trajectory in ascent, cruise, and descent that aims to minimize travel time. Our framework first fills the obstacle-free regions with intersecting convex polygons, then identifies potentially low-travel time candidate sequences of these polygons using a Graph of Convex Sets-based path planner, and then uses an Optimal Control Program to give the final trajectory that passes through the polygons in a sequence identified before. We evaluate our framework on routes within New York City. Our framework can design trajectories respecting the above constraints in the presence of as many as 250 building obstacles. We show that not including the above constraints can underestimate the flight time by as much as 20\%.
\end{abstract}


\section{Introduction} \label{Sec: introduction}
Increasing urbanization globally has overwhelmed existing transportation infrastructure, leading to traffic congestion and travel time delays. 
One proposed solution to reduce ground traffic congestion is Urban Air Mobility (UAM), which uses small-sized, high-speed electric aircraft capable of vertical takeoff and landing (called `eVTOLs') for intra-urban commute.  The successful adoption of UAM as an alternative means of transportation requires overcoming several operational challenges.

One such operational challenge is the design of a four-dimensional (time, latitude, longitude, altitude) trajectory that an eVTOL should follow while flying from one location to the other. 
Trajectory design is needed by eVTOL operators for estimating the eVTOL travel time and its energy consumption, and by urban planning bodies to identify suitable air corridors for eVTOL flight. As an eVTOL is expected to fly at low altitudes in dense urban environments, a rigorous trajectory design should account for obstacles arising from in-path buildings.
It should also respect real-world constraints imposed by wind, eVTOL dynamics, passenger comfort and safety, actuator bounds, and battery energy considerations in different flight phases. In this paper, we address the following question: How can we design trajectories for UAM operations that i) aim to minimize travel time, ii) account for dense building obstacles, and iii) respect complex eVTOL flight dynamics and other operational constraints in ascent, cruise, and descent in the presence of wind? 

\subsection{Related Work} \label{Subsec: related work}
Trajectory design for UAM has been investigated through different methods, including graph-based planning, traditional optimal control-based trajectory generation, and hybrid frameworks that combine global planning with local optimization. 

\noindent \textbf{Graph-based techniques}: These methods are popular due to their conceptual simplicity and strong capability to represent complex obstacle geometries. The environment is discretized into a grid, and shortest path algorithms such as the Dijkstra's or improved A* variants are applied to compute collision-free, optimal paths with respect to distance, time, or other cost metrics \cite{Dai2021, Blasi2020, Chati2025, Ye2024}. However, most graph-based planners operate at a kinematic level and do not explicitly model vehicle dynamics, actuator limits, or energy consumption. Although some works introduce simplified dynamics constraints or post-processing steps, these are approximations and do not guarantee feasibility of dynamics for eVTOLs in multiple flight phases \cite{Marcucci2023}. 


\noindent \textbf{Optimal control-based approaches}: In contrast, traditional optimal control-based approaches explicitly incorporate vehicle dynamics, operational constraints, and performance objectives within a single optimization framework. A large body of work formulates trajectory planning as an Optimal Control Program (OCP) and solves it using indirect methods \cite{Mall2024}, direct methods \cite{Pradeep2020, Lu2023}, or approximate convex programming techniques \cite{Wang2021, Wu2024}. These approaches can produce dynamically feasible trajectories and directly optimize metrics such as time or energy. However, many such existing studies neglect obstacle constraints in the optimization completely. Directly embedding such constraints in the OCP often yields highly nonconvex problems, making convergence difficult and limiting scalability in dense environments with many obstacles. Moreover, many works treat the different flight phases separately, and do not address the full eVTOL mission as a single unified optimization problem.

\noindent \textbf{Hybrid approaches}: Hybrid approaches combine the complementary strengths of graph-based and optimal control-based  approaches. Typically, a graph/grid-based planner first searches for a collision-free path or a sequence of safe regions, which then serves as guidance for a subsequent trajectory optimization stage \cite{Sun2025, Natarajan2021, Natarajan2024, Marcucci2023}. Such hybrid methods can improve robustness in cluttered environments while preserving dynamics feasibility.  However, current hybrid frameworks do not tackle end-to-end multi-phase eVTOL trajectory design as a unified optimization problem. They also neglect the effect of wind. 

Overall, the literature reveals a clear gap: there is a need for trajectory design methods that simultaneously handle obstacles, nonlinear vehicle dynamics in the presence of wind, operational constraints, and multiple flight phases within a unified and scalable framework. Not considering these factors can incorrectly estimate the eVTOL time of flight and energy consumption, which can lead to sub-optimal air corridors \cite{Chati2025}.

\subsection{Our Contributions} \label{Subsec: contributions}
We develop a unified hybrid framework that can integrate obstacles, wind, eVTOL flight dynamics, and other operational constraints to design a flight trajectory in vertical ascent, level cruise, and vertical descent that seeks to minimize the end-to-end flight time. We consider obstacles only in cruise. Instead of optimizing the trajectory by directly making it avoid the obstacles (which is a problem with nonconvex constraints that does not scale well with the number of obstacles), our framework identifies the complementary obstacle-free regions, fills them with intersecting convex polygons, and then optimizes the trajectory by enforcing it to pass through these polygons (resulting in a more scalable problem with convex obstacle constraints). 
\begin{figure}[b]
	\centering
	\includegraphics[width=0.5\columnwidth, trim=0cm 0cm 0cm 0cm, 
	clip]{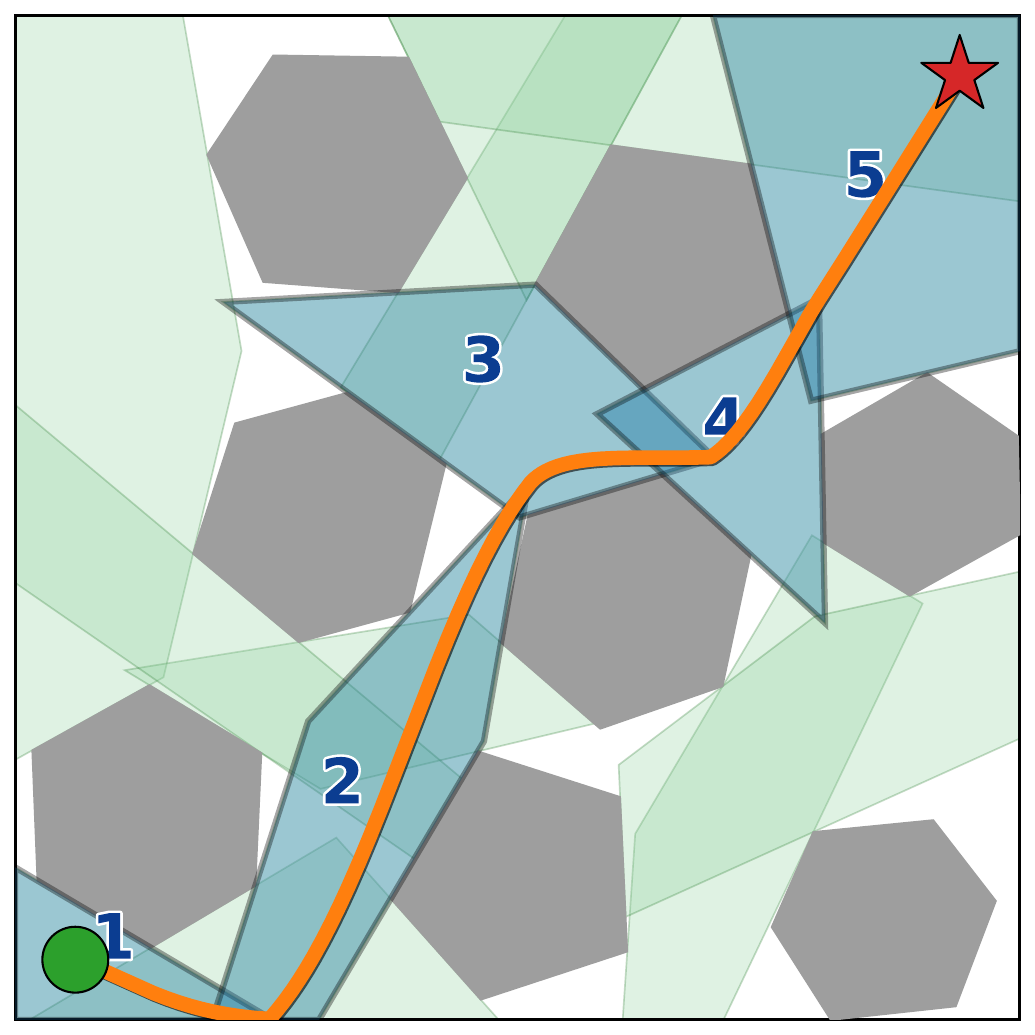}
	\caption{\label{Fig: schematic} Schematic of the GCS-OCP framework in cruise.}
\end{figure}

Fig. \ref{Fig: schematic} shows a schematic of the framework in cruise for ease of understanding. First, intersecting convex polygons (green and blue polygons in Fig. \ref{Fig: schematic}) are generated in the obstacle (gray polygons in Fig. \ref{Fig: schematic})-free space using the Iterative Regional Inflation by Semi-definite programming (IRIS) algorithm \cite{Deits2014}. 
Next, by creating a graph on these polygons, a Graph of Convex Sets (GCS)-based planner \cite{Marcucci2024, Marcucci2023} identifies probable candidate sequences of these polygons that could give trajectories with low travel time. In Fig. \ref{Fig: schematic}, one such candidate sequence is shown by the blue polygons ordered as 1 $->$ 2 $->$ 3 $->$ 4 $->$ 5. Lastly, an OCP gives the final shortest-time obstacle-free trajectory by enforcing it to pass through the sequence of the polygons identified by the GCS-based planner, and satisfy the nonlinear eVTOL flight dynamics in the presence of wind and other operational constraints 
(like maximum battery power and energy consumed, vehicle speed and acceleration limits, angular orientation and rate limits). In Fig. \ref{Fig: schematic}, such a trajectory connecting the start point (green circle) with the end point (red star) is shown in orange. It passes through the polygons in sequence 1 $->$ 2 $->$ 3 $->$ 4 $->$ 5. In the OCP, the first part of the orange cruise trajectory is constrained to lie in polygon 1, the second part in polygon 2, and so on.
Though GCS-based planning algorithms exist in prior literature \cite{Marcucci2023}, they use simplistic methods (like B\'ezier curves) to get the final trajectory through the sequence of safe regions identified. It is difficult to enforce the complex eVTOL flight dynamics constraints on such curves. \textit{Our novel contribution in this paper arises from merging OCP (with its ability to formulate any complex dynamics and constraints) with the existing GCS-based planner (with its ability to find a high-level path through obstacle-free regions) to optimize for the final trajectory.}
This GCS-OCP hybrid framework estimates not only the flight path coordinates but also the heading, airspeed, acceleration, thrust, pitch, roll, power, and energy consumption needed to achieve the trajectory in ascent, cruise, and descent.  
Based on the analyses done in this paper for New York City, a few specific findings are:
\begin{itemize}
	\item Not considering realistic constraints imposed by wind, obstacles, and flight dynamics can underestimate the end-to-end flight time by as much as 20\%.
	\item By enforcing the trajectory to \textit{pass} through obstacle-free regions, our GCS-OCP hybrid framework outputs trajectories with lower travel time than an alternative Direct OCP-only (D-OCP) formulation that seeks to actively \textit{avoid} obstacles.
	\item Our GCS-OCP framework scales with the number of obstacles better than D-OCP, and can give trajectories with as many as 250 obstacles (that D-OCP cannot).
\end{itemize}

\subsection{Outline} \label{Subsec: outline}
Section \ref{Sec: framework} describes the OCP formulation for eVTOL trajectory design. Section \ref{Sec: GCS OCP} gives details of the GCS-OCP hybrid framework developed in this paper and Sec. \ref{Sec: implementation} describes how it is implemented. Section \ref{Sec: results}
discusses important results and some limitations of our work. Section \ref{Sec: conclusions} concludes.  
\section{OCP for eVTOL Trajectory Design} \label{Sec: framework} Trajectory planning using OCP seeks to obtain a path (trajectory) for the state ($\mathbf{s}(t) \in \mathbb{R}^n$) of a dynamical system to go from a prescribed start point to a prescribed end point over time ($t \in \mathbb{R}^+\cup\{0\}$). The trajectory is chosen to minimize a desired objective/performance index ($\mathscr{O} \in \mathbb{R}$). The trajectory should also respect various state dynamics ($\mathscr{f}$), path ($\mathscr{C}$), and boundary ($\mathscr{B}$) constraints. Such a trajectory is realized by applying control inputs ($\mathbf{u}(t) \in \mathbb{R}^m$) to the system over time.  

We consider our system to be a quadcopter eVTOL with no lifting surfaces based on \cite{Pradeep2020}, where the dominant forces are rotor thrust, weight, and aerodynamic drag. The trajectory comprises three phases: (i) vertical ascent from the origin with (time, latitude, longitude, altitude) coordinates $(t_0=0,\lambda_0,\tau_0,h_0)$, (ii) level cruise at altitude $h_c$, and (iii) vertical descent to the destination with coordinates $(t_F,\lambda_F,\tau_F,h_F)$. For simplicity, we assume a constant wind field $\mathbf{W}:=(W_E,W_N)$, where $W_E$ and $W_N$ are the eastward and northward components, respectively. The wind field does not vary with time or position and has no vertical component. 

\subsection{States, Controls, Derived Variables, and Objective} \label{Subsec: variables} The system states $\mathbf{s}$ represent the vehicle position and motion, whereas controls $\mathbf{u}$ correspond to actuator inputs. For ease of OCP convergence, we assume different state and control variables for the different flight phases as follows:
\begin{itemize} 
	\item \textbf{Ascent/Descent:} The state vector is  $\mathbf{s}:=[h,V_v]^\top$, where $h$ is the altitude Above Mean Sea Level (AMSL) and $V_v$ is the vertical airspeed. The control input is $\mathbf{u}:=[T,\theta]^\top$, where $T$ is the total rotor thrust and $\theta$ is the rotor pitch. \item \textbf{Cruise:} The state vector is $\mathbf{s}:=[\lambda,\tau,h,\psi,V_l]^\top$, where $(\lambda,\tau)$ denotes latitude and longitude, $\psi$ is the heading, and $V_l$ is the horizontal airspeed. The control input vector is $\mathbf{u}:=[T,\theta,\phi]^\top$, where $\phi$ is the eVTOL roll angle. 
\end{itemize} 

We use the state and control variables to derive the eVTOL airspeed $V$, aerodynamic drag $D$, rotor angle of attack $\alpha$, and battery power $P$ as follows \cite{Pradeep2020, Johnson1994}: 
\begin{align} 
	V &= \sqrt{V_l^2 + V_v^2}, \nonumber \\ D &= 0.5\rho V^2\mathcal{S}C_D, \nonumber \\ \alpha &= \theta + \tan^{-1}\left(\frac{V_v}{V_l}\right), \nonumber \\ P &= \frac{\kappa T v_i + 0.125\rho\pi\Omega^3R^5\sigma C_{d_r}F_p + \max(TV\sin\alpha,0)}{\eta}. \label{Eq: derived variables} 
	\end{align} 
	Here, $\rho$ is the air density (= 1.225 kgm$^{-3}$), $\mathcal{S}$ is the eVTOL reference area, and $C_D$ is the eVTOL drag coefficient. The induced velocity $v_i$ from rotor downwash is given as the real, positive solution of 
	\begin{equation} v_i = \frac{v_h^2} {\sqrt{(V\cos\alpha)^2 + (V\sin\alpha+v_i)^2}}, \label{Eq: induced_velocity} 
	\end{equation} where $v_h$ is the induced velocity in hover and is given by \begin{equation} v_h = \sqrt{\frac{T}{2n_r\rho\pi R^2}}. \end{equation} The profile power factor is 
	\begin{equation} F_p = 1 + 4.6\left(\frac{V\cos\alpha}{\Omega R}\right)^2. \end{equation} 
	
	In the above equations, $\kappa$ is the induced power factor, $\Omega$ is the rotor angular speed, $R$ is the rotor radius, $\sigma$ is the rotor solidity ratio, $C_{d_r}$ is the rotor mean blade drag coefficient, $n_r$ is the number of rotors, and $\eta$ is an efficiency parameter. The battery energy consumed till time $t$ is 
	\begin{equation} 
		E(t) = \int_0^t P\,dt. 
	\end{equation} 
	The objective $\mathscr{O}$ to be minimized is the total travel time $t_F$. 
	
\subsection{Flight Dynamics Constraints} \label{Subsec: dynamics} The trajectory must satisfy nonlinear eVTOL flight dynamics $\mathscr{f}$ \cite{Pradeep2020}, \cite{Johnson1994} in all three phases. In ascent and descent, the dynamics are 
\begin{align} \frac{dh}{dt} &= V_v, \nonumber \\ \frac{dV_v}{dt} &= \frac{T\cos\theta - D\sin\gamma}{m} - g. \label{Eq: ascent_descent_dynamics} \end{align} 
In cruise, the dynamics are 
\begin{align} \frac{d\lambda}{dt} &= \frac{V_l\cos\psi + W_N}{R_e+h}, \nonumber \\ \frac{d\tau}{dt} &= \frac{V_l\sin\psi + W_E}{(R_e+h)\cos\lambda}, \nonumber \\ \frac{dh}{dt} &= 0, \nonumber \\ \frac{d\psi}{dt} &= \frac{T\sin\phi}{mV_l}, \nonumber \\ \frac{dV_l}{dt} &= \frac{T\cos\phi\sin\theta - D\cos\gamma}{m}. \label{Eq: cruise_dynamics} \end{align} 
Here, $m$ is the eVTOL mass, $g$ is the acceleration due to gravity  (= 9.8 ms$^{-2}$), $R_e$ is the Earth's radius (= 6,371 km), and $\gamma=\tan^{-1}(V_v/V_l)$ is the flight path angle. 
	
\subsection{Path and Boundary Constraints} \label{Subsec: path constraints} 
The trajectory must also satisfy operational constraints that represent actuator and battery limits and ensure passenger safety and comfort. A knowledge of the eVTOL statics and dynamics gives the following phase-specific path constraints $\mathscr{C}$: 
\begin{itemize} \item \textbf{Ascent/Descent:} \begin{equation} \begin{array}{l} V_l = \sqrt{W_E^2+W_N^2}, \\ \phi = 0, \quad \psi = \tan^{-1}\left(\frac{-W_E}{-W_N}\right), \\ T\sin\theta = D\cos\gamma. \end{array} \end{equation} \item \textbf{Cruise:} $V_v = 0$, $T\cos\phi\cos\theta = mg$. \end{itemize} 
We also impose bounds on vertical airspeed, horizontal airspeed, pitch, roll, pitch rate, roll rate, horizontal acceleration, vertical acceleration, thrust, power, and battery energy consumed: \begin{align} V_v &\in [0,V_{v_{\max}}] \quad \textrm{in ascent}, \nonumber \\ V_v &\in [-V_{v_{\max}},0] \quad \textrm{in descent}, \nonumber \\ V_l &\in [0,V_{l_{\max}}], \nonumber \\ |\theta| &\leq \theta_{\max}, \quad |\phi| \leq \phi_{\max}, \nonumber \\ |\dot{\theta}| &\leq \dot{\theta}_{\max}, \quad |\dot{\phi}| \leq \dot{\phi}_{\max}, \nonumber \\ |a_l| &\leq a_{\max}, \quad |a_v| \leq a_{\max}, \nonumber \\ T &\geq 0, \quad P \in [0,P_{\max}], \quad E \in [0,E_{\max}]. \label{Eq: operational_bounds} \end{align} 
We enforce continuity of states across phase transitions. The boundary conditions $\mathscr{B}$ are \begin{equation} h(t_0)=h_0, \quad h(t_F)=h_F, \quad V_v(t_0)=V_v(t_F)=0. \label{Eq: boundary_conditions} \end{equation} 

\subsection{Obstacle Avoidance Constraint in Direct-OCP} \label{Subsec: obstacles direct OCP} We incorporate obstacles during the cruise phase only. Obstacles arise from buildings whose heights ${h_o}_i$ plus a vertical safety buffer $L_v$ exceed the cruising altitude $h_c$. We also apply a lateral buffer $L_l$ around each building planform. In the Direct-OCP (D-OCP) method, each obstacle is approximated as a circle in the horizontal cruise plane. Its center, with coordinates $({\lambda_o}_i,{\tau_o}_i)$, lies at the centroid of the buffered building planform, and its radius ${r_o}_i$ equals the maximum distance from the planform centroid to its vertices. This circular approximation is used only for ease of constraint formulation in D-OCP. Let \begin{equation} O := \{({\lambda_o}_i,{\tau_o}_i,{r_o}_i,{h_o}_i) \mid {h_o}_i + L_v > h_c\} \end{equation} denote the set of all $n_o$ obstacles. Obstacle avoidance requires each point $(\lambda,\tau)$ on the cruise trajectory to lie outside all obstacle circles: \begin{equation} G((\lambda,\tau),({\lambda_o}_i,{\tau_o}_i)) \geq {r_o}_i \quad \forall i \in \{1,2,\ldots,n_o\}, \label{Eq: obstacle constraint} \end{equation} where $G(x,y)$ denotes the great-circle distance between $x$ and $y$. This constraint is nonconvex and therefore, D-OCP faces convergence issues when the number of obstacles is large.
\section{Hybrid Framework for Trajectory Design} \label{Sec: GCS OCP} To address the problem of nonconvexity of obstacle constraints in D-OCP, we reformulate obstacle handling by constraining the trajectory to \textit{pass} through a sequence of intersecting convex regions in the obstacle-free space instead of enforcing it to \textit{avoid} obstacles. Thus, we develop a hybrid framework called `GCS-OCP' in Algorithm~\ref{Alg}. Given start ($\mathscr{s}$) and end ($\mathscr{g}$) points and a cruising altitude ($h_c$), GCS-OCP integrates convex decomposition, graph-based search, and optimal control to compute the final trajectory as described below. \begin{algorithm}[tbp]
	\DontPrintSemicolon
	\SetKwInput{Inputs}{\underline{Inputs}}
	\SetKwInput{Trajectory}{\underline{Trajectory design}}
	
	\Inputs{}
	$\mathscr{s}, \mathscr{g}, h_c $ \tcp*{Prescribed conditions} 
	
	$p_0 := (\lambda_0, \tau_0, h_c)$ \tcp*{Start of cruise} 
	
	$p_J := (\lambda_F, \tau_F, h_c)$ \tcp*{End of cruise}
	
	BD, $\bm{W}$  \tcp*{Buildings and wind datasets} 
	
	$\mathbf{s}, \mathbf{u}, \mathscr{O}, \mathscr{f}, \mathscr{C}, \mathscr{B}, N_\textrm{cr}$ \tcp*{OCP quantities}
	
	
	\Trajectory{} 
	$J$ grid points $\bm{p}$ $\gets$ \textproc{CreateGrid}($p_0, p_J$) \label{Alg: grid}\;
	
	$\mathcal{O}$ $\gets$ \textproc{GetObstacles}(BD, $L_v$, $L_l$, $h_c$)\label{Alg: obstacles}\;
	
	$\mathcal{P}$ $\gets$ \textproc{GetConvexPolygons}($\bm{p}, \mathcal{O}$) \tcp*{IRIS} \label{Alg: IRIS}
	
	$\mathcal{G}$ $\gets$ \textproc{CreateGcs}($\mathcal{P}, \bm{W}$) \label{Alg: GCS graph}
	
	$\mathcal{C}$ $\gets$ \textproc{GetCandidatePaths}($\mathcal{G}, \mathcal{P}, p_0, p_J$) \tcp*{GCS-Planner} \label{Alg: GCS path} 
	
	\For(\tcp*[f]{Candidate paths}){$\epsilon \in \mathcal{C}$} {
		\For(\tcp*[f]{Cruise intervals}){$N_\textrm{cr} \in [20, 100]$} {
			
			$\textrm{Traj}(\epsilon, N_\textrm{cr}), t_F(\epsilon, N_\textrm{cr})$ $\gets$  \tcp*{OCP} \textproc{RunOcp}($\mathscr{s}, \mathscr{g}, h_c, \bm{W}, \mathbf{s}, \mathbf{u}, \mathscr{O}, \mathscr{f}, \mathscr{C}, \mathscr{B}, \mathcal{P}, \epsilon, N_\textrm{cr}$) \label{Alg: GCS-OCP} 
		}
	}
	$\epsilon^*, N_\textrm{cr}^*$ $\gets$ $\underset{\epsilon \in \mathcal{C}, N_\textrm{cr}}{\mathrm{arg min}}\,t_F(\epsilon, N_\textrm{cr})$
	
	$\textrm{Traj}^*$ $\gets$ $\textrm{Traj}(\epsilon^*, N_\textrm{cr}^*)$;
	$t_F^*$ $\gets$ $t_F(\epsilon^*, N_\textrm{cr}^*)$ \tcp*{Output}
	
	\caption{Algorithm for GCS-OCP framework}
	\label{Alg}
\end{algorithm} 

\subsection{Convex Decomposition of Obstacle-Free Space} \textproc{CreateGrid} (line~\ref{Alg: grid} in Algorithm~\ref{Alg}) first generates a two-dimensional horizontal grid of $J$ grid points $\mathbf{p}$ in the cruise plane. These grid points are placed along and perpendicular to the straight-line path from the start of cruise, $p_0 := (\lambda_0,\tau_0,h_c)$ (green circle in Fig. \ref{Fig: schematic}), to the end of cruise, $p_J := (\lambda_F,\tau_F,h_c)$ (red star in Fig. \ref{Fig: schematic}). Each point is at most 141 m away from its immediate neighbor. The grid extends to some half-width (500--1500 m, depending on the scenario) on either side of the straight-line path. We conduct all geometric computations in the New York City (NYC)-specific NAD83 Cartesian coordinate system.

Using the buildings' dataset BD, \textproc{GetObstacles} (line~\ref{Alg: obstacles}) generates the set of polygonal building obstacles $\mathcal{O}$ (gray polygons in Fig. \ref{Fig: schematic}) in the cruise plane. As described in Sec.~\ref{Subsec: obstacles direct OCP}, we include only those buildings whose height plus the vertical safety buffer $L_v$ exceeds the cruising altitude $h_c$. We also apply a lateral safety buffer $L_l$ around each such building planform. We retain only those obstacles whose buffered planform polygons intersect the rectangular grid generated above. The obstacles appear as two-dimensional polygons in this cruise grid plane when viewed from above. In contrast to D-OCP, where obstacles are approximated as circles for ease of constraint formulation, GCS-OCP uses the actual polygonal planforms as obstacles.

\textproc{GetConvexPolygons} (line~\ref{Alg: IRIS}) uses the Iterative Regional Inflation by Semi-definite programming (IRIS) algorithm~\cite{Deits2014} to decompose the obstacle-free region into intersecting convex polygons (green and blue polygons in Fig. \ref{Fig: schematic}). Using each grid point as a seed, IRIS starts by creating a small initial ellipse centered at the seed point. It then constructs separating hyperplanes between this ellipse and nearby obstacles. The intersection of these separating hyperplanes and the sides of the grid bounding rectangle gives a convex polygon enclosing the ellipse that lies in the obstacle-free region. IRIS then finds a larger ellipse inside this convex polygon and repeats the process until the growth of the ellipse becomes sufficiently small. On converging, the algorithm returns the final obstacle-free
convex polygon and the enclosed ellipse for each
seed point. Running IRIS for all the seed points results in a collection of $n_p$ intersecting convex polygons $\mathcal{P}$ that cover the obstacle-free region. We discard polygons that lie completely inside another polygon or are near-duplicates of already generated polygons. \begingroup \setlength{\abovedisplayskip}{3pt} \setlength{\belowdisplayskip}{3pt} \setlength{\abovedisplayshortskip}{3pt} \setlength{\belowdisplayshortskip}{3pt} The ellipse generated in IRIS can be represented as the affine image of a unit ball: \begin{equation} 
	\mathscr{E}(\mathbf{C},\mathbf{d}) := \{\mathbf{x}=\mathbf{C}\tilde{\mathbf{x}}+\mathbf{d} \mid \|\tilde{\mathbf{x}}\|_2 \leq 1\}. \label{Eq: iris_ellipse} \end{equation} 
	\endgroup Here, $\mathbf{C}\in\mathbb{R}^{2\times 2}$ defines the shape of the ellipse $\mathscr{E}$ and $\mathbf{d}\in\mathbb{R}^2$ defines its center. After convergence, each obstacle-free convex polygon $\mathcal{P}_i$ can be represented in half-space form as \begingroup \setlength{\abovedisplayskip}{3pt} \setlength{\belowdisplayskip}{3pt} \setlength{\abovedisplayshortskip}{3pt} \setlength{\belowdisplayshortskip}{3pt} \begin{equation} \mathcal{P}_i := \{\mathbf{x}\in\mathbb{R}^2 \mid \mathbf{A}_i\mathbf{x} \leq \mathbf{b}_i\}. \label{Eq: convex_polygon} \end{equation} \endgroup Here, $\mathbf{x} \in \mathbb{R}^2$ are the (Easting, Northing) NAD83-Cartesian coordinates, $\mathbf{A}_i\in\mathbb{R}^{\zeta_i\times 2}$ and $\mathbf{b}_i\in\mathbb{R}^{\zeta_i}$ define the $\zeta_i$ half-space constraints of polygon $\mathcal{P}_i$. We refer the reader to~\cite{Deits2014} for the mathematical details of the optimization problem solved in IRIS. 

\subsection{Graph-Based Search Using GCS} \textproc{CreateGcs} (line~\ref{Alg: GCS graph}) constructs a bidirectional Graph of Convex Sets (GCS) $\mathcal{G}:=(\mathcal{V},\mathcal{E})$. Each convex polygon $\mathcal{P}_i$ is a vertex $\nu_i \in \mathcal{V}$ of the graph. We create an edge $e_{ij}\in\mathcal{E}$ from vertex $\nu_i$ to vertex $\nu_j$ if the corresponding polygons $\mathcal{P}_i$ and $\mathcal{P}_j$ intersect. Each edge is weighted by a function that gives the time taken to travel, in the presence of wind, between any two points in the two different intersecting polygons that are connected by the edge. If edge $e_{ij}$ has a starting point $\mathbf{x}_e \in \mathcal{P}_i$, and an end point $\mathbf{x}'_e \in \mathcal{P}_j$, then we define this edge weight function as \begingroup \setlength{\abovedisplayskip}{3pt} \setlength{\belowdisplayskip}{3pt} \setlength{\abovedisplayshortskip}{3pt} \setlength{\belowdisplayshortskip}{3pt} \begin{equation} w_e := \frac{\|\mathbf{x}'_e-\mathbf{x}_e\|_2}{V_{g,e}}, \label{Eq: edge_weight} \end{equation} \endgroup where $V_{g,e}$ is the estimated ground speed along the directed edge. We estimate $V_{g,e}$ by combining the horizontal airspeed vector and the wind velocity vector $\mathbf{W}$ along the edge direction. Thus, the GCS provides a high-level graph representation of the obstacle-free cruise region and approximately accounts for the effect of wind through the edge weights. 

\textproc{GetCandidatePaths} (line~\ref{Alg: GCS path}) then uses a GCS-based planner~\cite{Marcucci2023, Marcucci2024} to generate a small set $\mathcal{C}$ of candidate polygon sequences $\epsilon$. These polygon sequences connect the polygon containing $p_0$ to the polygon containing $p_J$ and can potentially give low-travel-time cruise trajectories.  The GCS-based planner associates a flow variable $y_e$ with each edge $e$. In the exact formulation, $y_e\in\{0,1\}$ indicates whether edge $e$ is selected as part of a polygon sequence. We also define an auxiliary variable $\beta_e\geq 0$ for the objective and two perspective variables \begingroup \setlength{\abovedisplayskip}{3pt} \setlength{\belowdisplayskip}{3pt} \setlength{\abovedisplayshortskip}{3pt} \setlength{\belowdisplayshortskip}{3pt} \begin{equation} \mathbf{z}_e := y_e\mathbf{x}_e, \qquad \mathbf{z}'_e := y_e\mathbf{x}'_e. \label{Eq: auxiliary_variables} \end{equation} \endgroup

The GCS-based polygon sequence-selection problem can then be written as a mixed-integer convex program. The objective minimizes the sum of auxiliary edge-cost variables: \begin{equation} \min_{\{\beta_e,y_e,\mathbf{z}_e,\mathbf{z}'_e\}} \sum_{e\in\mathcal{E}}\beta_e . \label{Eq: GCS objective} \end{equation} The edge-cost constraint is \begin{equation} V_{g,e}\beta_e \geq \|\mathbf{z}'_e-\mathbf{z}_e\|_2, \quad \forall e\in\mathcal{E}. \label{Eq: GCS edge cost} \end{equation} The source and sink flow constraints are \begin{equation} \sum_{e\in\mathcal{E}^{\textrm{out}}_{p_0}} y_e = 1, \quad \sum_{e\in\mathcal{E}^{\textrm{in}}_{p_J}} y_e = 1. \label{Eq: GCS source sink} \end{equation} For each intermediate vertex, flow conservation and the no-revisit constraint are imposed as \begin{align} \sum_{e\in\mathcal{E}^{\textrm{out}}_{\nu}} y_e &= \sum_{e\in\mathcal{E}^{\textrm{in}}_{\nu}} y_e, && \forall \nu\in\mathcal{V}\setminus\{p_0,p_J\}, \label{Eq: GCS flow conservation} \\ \sum_{e\in\mathcal{E}^{\textrm{out}}_{\nu}} y_e &\leq 1, && \forall \nu\in\mathcal{V}\setminus\{p_0,p_J\}. \label{Eq: GCS no revisit} \end{align} Continuity of the selected sequence in the convex sets is enforced through \begin{equation} \sum_{e\in\mathcal{E}^{\textrm{out}}_{\nu}} \mathbf{z}_e = \sum_{e\in\mathcal{E}^{\textrm{in}}_{\nu}} \mathbf{z}'_e, \quad \forall \nu\in\mathcal{V}\setminus\{p_0,p_J\}. \label{Eq: GCS point continuity} \end{equation} For each edge $e_{ij}$, the edge points are constrained to lie inside the corresponding convex polygons: \begin{equation} \mathbf{A}_i\mathbf{z}_e \leq \mathbf{b}_i y_e, \quad \mathbf{A}_j\mathbf{z}'_e \leq \mathbf{b}_j y_e, \quad \forall e_{ij}\in\mathcal{E}. \label{Eq: GCS set membership} \end{equation} Finally, the exact sequence-selection formulation requires \begin{equation} y_e \in \{0,1\}, \quad \forall e\in\mathcal{E}. \label{Eq: GCS binary} \end{equation} Here, $V_{g,e}$ is the estimated ground speed along edge $e$, and $\mathcal{E}^{\textrm{out}}_{\nu}$ and $\mathcal{E}^{\textrm{in}}_{\nu}$ denote the sets of outgoing and incoming edges at vertex $\nu$, respectively.
To reduce computational cost, we solve a convex relaxation of the above mixed-integer formulation in which $y_e$ is allowed to take any real value in $[0,1]$ instead of being constrained to be binary. Such relaxations are tight in practice for many GCS planning problems~\cite{Marcucci2023}. Instead of using an exact branch-and-bound solution technique, we solve this convex relaxation fast to obtain an approximate solution for the flow variables and edge points. Based on the relaxed solution, we generate a candidate sequence $\epsilon$ by selecting outgoing edges with probabilities proportional to their flow-variable values, while ensuring that no vertex is visited more than once. We repeat this randomized procedure multiple times to obtain multiple candidate polygon sequences. We also include a greedy candidate sequence obtained by repeatedly choosing the outgoing edge with the largest flow value. We then order the candidate sequences by their estimated travel time, computed as the sum of the edge weights $w_e$ along each sequence, and retain only the top few candidates to give a small set $\mathcal{C}$ of polygon sequences $\epsilon$. We refer the reader to \cite{Marcucci2024, Marcucci2023} for more mathematical details of the GCS-based planner. (In Fig. \ref{Fig: schematic}, the blue polygons ordered from 1 to 5 give one such sequence.)

\subsection{Trajectory Design via Optimal Control} For each candidate sequence $\epsilon \in \mathcal{C}$, \textproc{RunOcp} (line~\ref{Alg: GCS-OCP}) runs an OCP to obtain the actual end-to-end trajectory in ascent, cruise, and descent. The OCP uses the same states, controls, objective, flight dynamics, path, and boundary constraints as described in Sec.~\ref{Sec: framework}. However, this OCP differs from D-OCP in
the manner in which the obstacle constraint is formulated
in cruise. We partition the cruise phase into multiple segments and map these segments, in order, to the successive convex polygons in the candidate sequence $\epsilon$. If a cruise segment is mapped to polygon $\mathcal{P}_i$, then each point $\mathbf{x}$ on that segment is constrained to satisfy \begingroup \setlength{\abovedisplayskip}{3pt} \setlength{\belowdisplayskip}{3pt} \setlength{\abovedisplayshortskip}{3pt} \setlength{\belowdisplayshortskip}{3pt} \begin{equation} \mathbf{A}_i\mathbf{x} \leq \mathbf{b}_i. \label{Eq: polygon_containment} \end{equation} \endgroup 
Therefore, we constrain the cruise trajectory to remain inside the selected obstacle-free convex polygons. This enforces obstacle avoidance using convex containment constraints, instead of directly requiring the trajectory to avoid each obstacle through nonconvex constraints as done in D-OCP. (In Fig. \ref{Fig: schematic}, such a trajectory is shown in orange, with its first part constrained to lie in blue polygon 1, the second part in polygon 2, and so on.) The use of convex polygon containment also changes the role of the graph-based planner. The GCS-planner is not used to directly output the final dynamically feasible trajectory. Instead, it identifies a small number of promising obstacle-free corridors. The OCP then searches within these corridors while enforcing the nonlinear flight dynamics and other operational constraints. This separation is useful because it avoids directly embedding many nonconvex obstacle-avoidance constraints in the OCP, while still allowing the final trajectory to be dynamically feasible. 

After running the OCP for each candidate sequence $\epsilon\in\mathcal{C}$, the final output trajectory (`Traj$^*$') is the one with the least total travel time ($t_F^*$) across all the trajectories obtained
for the different sequences in $\mathcal{C}$: 
\begingroup \setlength{\abovedisplayskip}{3pt} \setlength{\belowdisplayskip}{3pt} \setlength{\abovedisplayshortskip}{3pt} \setlength{\belowdisplayshortskip}{3pt} \begin{equation} \textrm{Traj}^* := \textrm{Traj}(\epsilon^*,N_{\textrm{cr}}^*); \qquad t_F^* := t_F(\epsilon^*,N_{\textrm{cr}}^*), \label{Eq: final_trajectory} \end{equation} \endgroup where
\begingroup \setlength{\abovedisplayskip}{3pt} \setlength{\belowdisplayskip}{3pt} \setlength{\abovedisplayshortskip}{3pt} \setlength{\belowdisplayshortskip}{3pt} \begin{equation} (\epsilon^*,N_{\textrm{cr}}^*) = \underset{\epsilon\in\mathcal{C},\,N_{\textrm{cr}}}{\arg\min} \; t_F(\epsilon,N_{\textrm{cr}}). \label{Eq: final_selection} \end{equation} \endgroup 

\section{Implementation} \label{Sec: implementation} 
In this section, we describe how we implement our trajectory
design framework.

\subsection{Geography, Input Datasets, and Parameters} \label{Subsec: geography} 
We evaluate our framework on routes in New York City (NYC) connecting different origin points to the John F. Kennedy Airport (JFK). We consider this use-case of commuting to an airport as it is projected to be one of the important initial use-cases for UAM adoption \cite{Rath2022}. We consider a range of cruising altitudes $h_c$ that give a varying number of building obstacles. These altitudes lie between a minimum permissible altitude $h_{\min}=136$ m AMSL and a maximum permissible altitude $h_{\max}=1225$ m AMSL. These altitude bounds correspond approximately to 400--4000 ft above NYC's mean ground elevation \cite{Qu2023, NYC_elevation}. We obtain building data from the NYC OpenData Building Footprints Database (BD) \cite{buildings}. We approximate a constant wind field $\mathbf{W}$ using the Open-Meteo Historical Weather API \cite{Open-Meteo}. This wind field has a constant northward component of wind $W_N=4.9$ ms$^{-1}$ and a constant eastward component $W_E=1.6$ ms$^{-1}$. Table \ref{Tab: parameters} shows all the operational parameters used in this paper. These parameters have been adopted from prior literature \cite{Pradeep2020, Chati2025, Johnson1994, Pradeep2021}. 

\begin{table}[t] \centering \caption{Key eVTOL and operational parameters used in analyses.} \label{Tab: parameters} \begin{tabular}{lll} \hline Parameter & Symbol & Value \\ \hline Maximum vertical airspeed & $V_{v_{\max}}$ & 1.5 ms$^{-1}$ \\ Maximum horizontal airspeed & $V_{l_{\max}}$ & 50.4 ms$^{-1}$ \\ Maximum pitch angle & $\theta_{\max}$ & $10^\circ$ \\ Maximum roll angle & $\phi_{\max}$ & $10^\circ$ \\ Maximum pitch rate & $\dot{\theta}_{\max}$ & $5^\circ$/s \\ Maximum roll rate & $\dot{\phi}_{\max}$ & $5^\circ$/s \\ Maximum acceleration & $a_{\max}$ & $g/\sqrt{2}$ \\ Maximum battery power & $P_{\max}$ & 494.25 kW \\ Reference area & $\mathcal{S}$ & 1.1984 m$^2$ \\ Drag coefficient & $C_D$ & 1 \\ Induced power factor & $\kappa$ & 1.15 \\ Rotor angular speed & $\Omega$ & 30.12 rad s$^{-1}$ \\ Rotor radius & $R$ & 4 m \\ Rotor solidity ratio & $\sigma$ & 0.055 \\ Mean blade drag coefficient & $C_{d_r}$ & 0.0089 \\ Number of rotors & $n_r$ & 4 \\ Efficiency parameter & $\eta$ & 0.9 \\ eVTOL mass & $m$ & 2940 kg \\ Vertical safety buffer & $L_v$ & 50 m \\ Lateral safety buffer & $L_l$ & 25 m \\ Usable battery energy & $E_{\max}$ & $0.85 \times 1331$ MJ \\ \hline \end{tabular} \end{table}

\subsection{OCP Discretization} \label{Subsec: transcription} 
We numerically solve the OCP by transcribing it into a Nonlinear Program (NLP) with discrete variables using direct (\textit{discretize then optimize}) methods \cite{Betts2010}. Such methods convert the original problem of optimization over infinite-dimensional function spaces to a more tractable optimization problem over finite-dimensional vector spaces of discrete variables.
We discretize the time window $[0,t_F]$ into $N=N_{\textrm{as}}+N_{\textrm{cr}}+N_{\textrm{de}}$ intervals in ascent, cruise, and descent, respectively, giving $N+1$ discrete time instants $0 = t_0, t_1, \ldots, t_k, t_{k+1}, \ldots, t_N = t_F$. We fix $N_{\textrm{as}}=10$ and $N_{\textrm{de}}=10$, and vary $N_{\textrm{cr}}$ between 20 and 100 to improve solver convergence. In the GCS-OCP framework, we further divide $N_{\textrm{cr}}$ equally among the different cruise trajectory segments that are mapped to successive polygons in the candidate polygon sequence generated by the GCS-planner. The time, state, and control variables at these $N+1$ discrete instants now serve as the decision variables for the NLP. For simplicity, we assume a straight-line path in space between two successive instants. We enforce constraints that ensure continuity of state and satisfaction of flight dynamics at each discrete
instant by applying the Forward Euler quadrature scheme to the dynamics functions $\mathscr{f}$. When solved, the NLP gives approximate estimates of the different state and control variables at the discrete time instants. 
\begin{figure*}[htb]
	\centering
	\begin{subfigure}[t]{0.35\textwidth}
		\centering
		\includegraphics[width=\textwidth, trim=0cm 0cm 0cm 0cm, 
		clip]{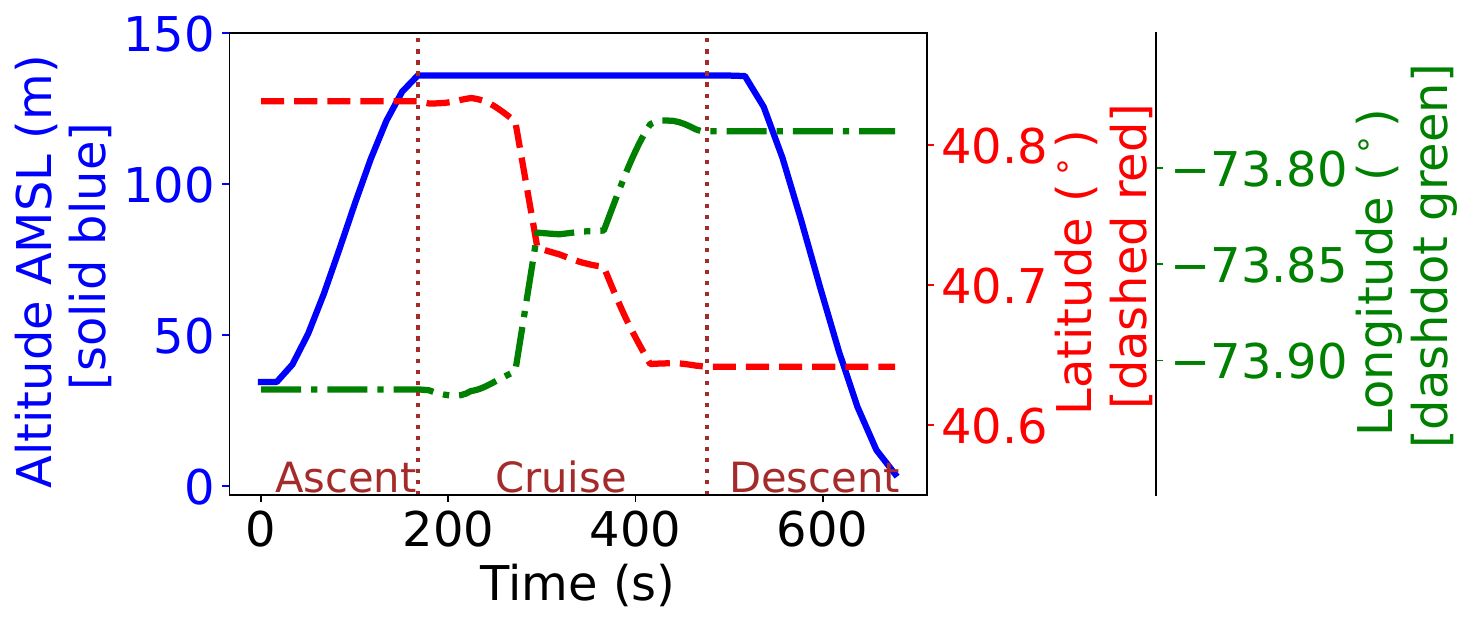}
		\caption{Altitude, latitude, and longitude.}
		\label{Fig: profiles}
	\end{subfigure}
	\begin{subfigure}[t]{0.64\textwidth}
		\centering
		\includegraphics[width=\textwidth, trim=0cm 5.5cm 2.5cm 2cm,
		clip]{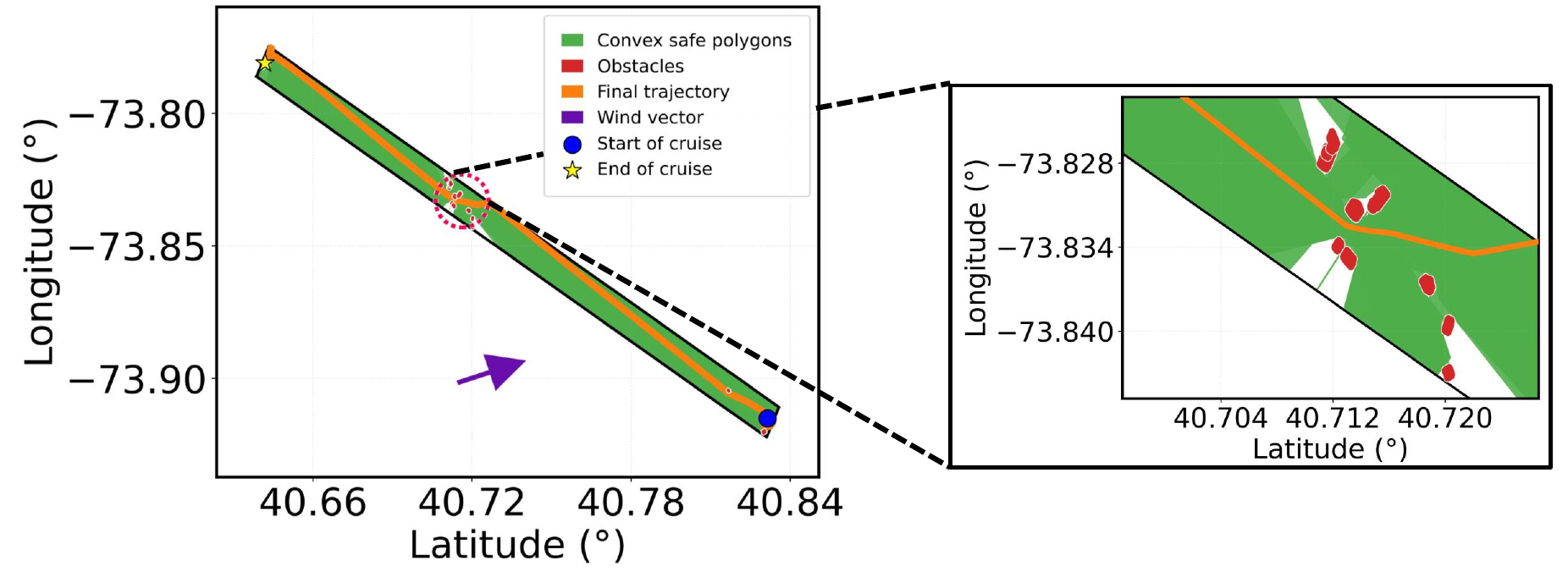}
		\caption{Obstacles and cruise trajectory (as seen from above).}
		\label{Fig: trajectory}
	\end{subfigure}
	
	\begin{subfigure}[t]{0.28\textwidth}
		\centering
		\includegraphics[width=\textwidth, trim=0cm 0cm 0cm 0cm, clip]{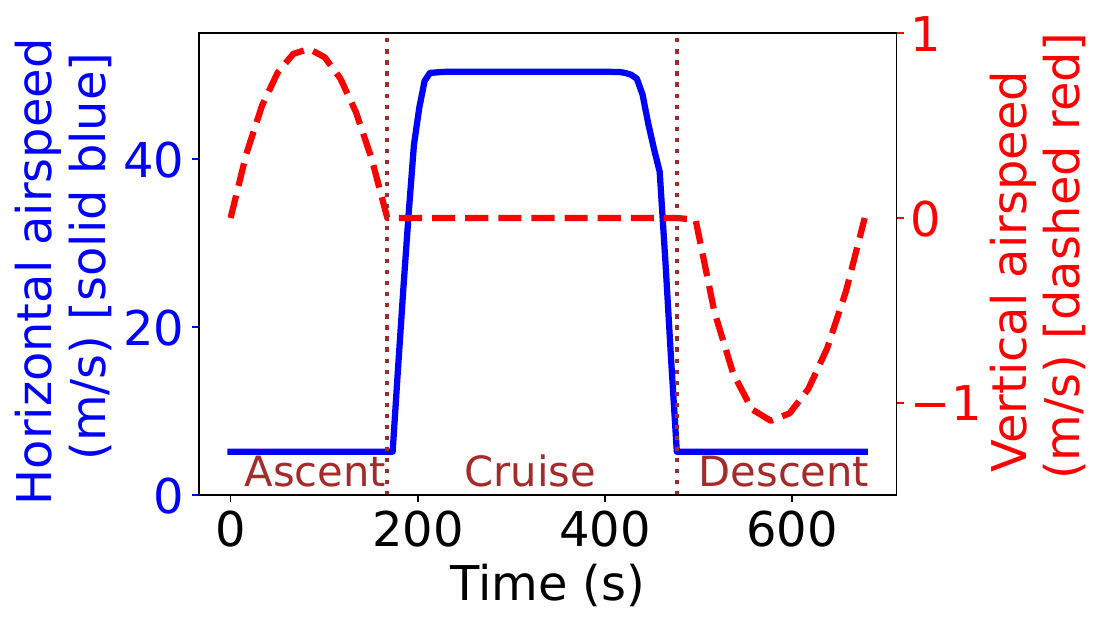}
		\caption{Airspeed.}
		\label{Fig: airspeed}
	\end{subfigure}
	\begin{subfigure}[t]{0.34\textwidth}
		\centering
		\includegraphics[width=\textwidth, trim=0cm 0cm 0cm 0cm, 
		clip]{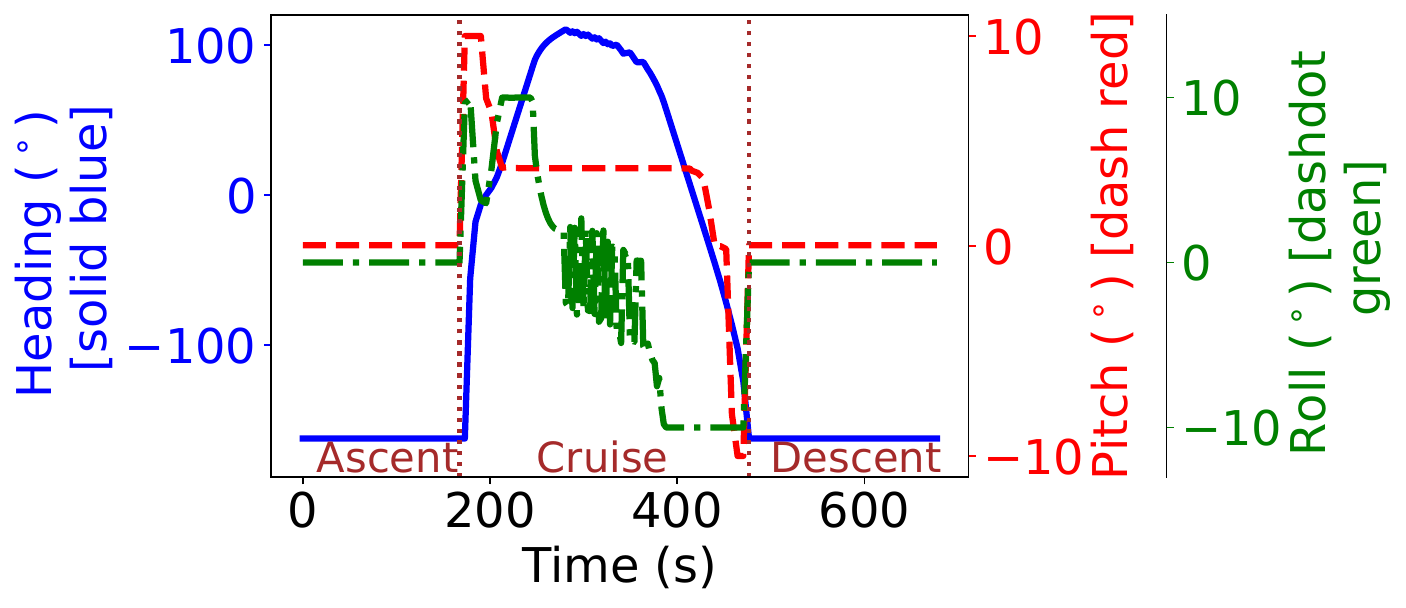}
		\caption{Heading, pitch, and roll.}
		\label{Fig: angles}
	\end{subfigure}
	\begin{subfigure}[t]{0.34\textwidth}
		\centering
		\includegraphics[width=\textwidth, trim=0cm 0cm 0cm 0cm, clip]{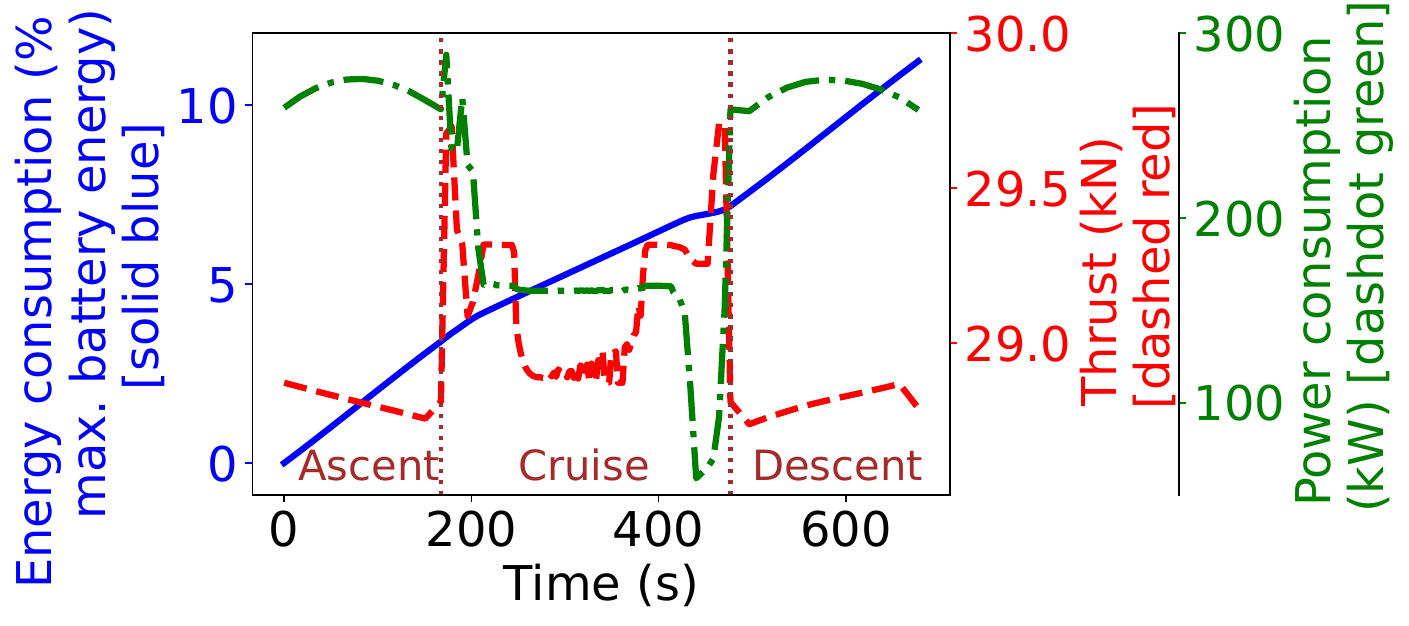}
		\caption{Energy, thrust, and power.}
		\label{Fig: thrust}
	\end{subfigure}
	
	\caption{\label{Fig: GCS output} Profiles of different output variables from GCS-OCP framework for one example flight to JFK. 
	}
\end{figure*}

\subsection{Obstacle Constraint Transcription} For D-OCP, we transcribe the nonconvex obstacle avoidance constraint for each of the $n_o$ circular obstacles in \eqref{Eq: obstacle constraint} as \begin{align} & G\left(\overline{(\lambda(t_k),\tau(t_k)):(\lambda(t_{k+1}),\tau(t_{k+1}))}, ({\lambda_o}_i,{\tau_o}_i)\right) \geq {r_o}_i \nonumber \\ & \hspace{0.4cm} \forall i \in \{1,\ldots,n_o\}, \quad \forall k \in \{N_{\textrm{as}},\ldots,N_{\textrm{as}}+N_{\textrm{cr}}-1\}. \label{Eq: obstacle transcribed D-OCP} \end{align} Here, $\overline{(\lambda(t_k),\tau(t_k)):(\lambda(t_{k+1}),\tau(t_{k+1}))}$ is the line segment between coordinates $(\lambda(t_k),\tau(t_k))$ and $(\lambda(t_{k+1}),\tau(t_{k+1}))$ at consecutive discrete time instants $t_k$ and $t_{k+1}$ in cruise. Thus, the distance between each such line segment and the center of each circular obstacle should be greater than or equal to the radius of the obstacle. By contrast, in the GCS-OCP formulation, obstacle avoidance is enforced through convex containment constraints. If a cruise trajectory segment is mapped to a polygon $\mathcal{P}_i$, then the discretized trajectory points for that segment should satisfy \begin{equation} \mathbf{A}_i\mathbf{x}(t_k) \leq \mathbf{b}_i \quad \forall k \in \{N_{\textrm{as}},\ldots,N_{\textrm{as}}+N_{\textrm{cr}}-1\}. \label{Eq: obstacle transcribed GCS-OCP} 
\end{equation} Here, $(\mathbf{A}_i,\mathbf{b}_i)$ define the obstacle-free convex polygon $\mathcal{P}_i$, and $\mathbf{x}(t_k)$ is the Cartesian projection of the cruise trajectory point at time instant $t_k$. Since the convex polygons may intersect, a trajectory point can lie in more than one polygon.

\subsection{Software Packages and Solvers} \label{Subsec: software} We perform all analyses in \texttt{Python3} on a desktop-class machine. We formulate the optimization problems involved in the IRIS algorithm and the GCS-based planner in the \texttt{CVXPY} modeling language for convex optimization problems \cite{Diamond2016, Agrawal2018}, and solve them using the \texttt{ECOS} open-source solver \cite{Domahidi2013}. We implement OCP using the \texttt{CasADi} open-source tool for nonlinear optimization \cite{Andersson2019}. \texttt{CasADi} numerically solves the transcribed NLP using the \texttt{IPOPT} solver \cite{Wachter2006}, with tolerance set to $10^{-4}$ and maximum number of iterations set to 3000.

\section{Results} \label{Sec: results}
\subsection{Output Trajectory and Other Variables} \label{Subsec: trajectory results}
Fig. \ref{Fig: GCS output} shows the profiles of the different variables output by our GCS-OCP framework for an example flight. The origin point in NYC is 24 km away from the destination JFK airport. We choose an altitude of 136 m which results in 13 obstacles. The bounds imposed on the accelerations result in smooth profiles for the positions and airspeeds, especially at the phase-change instants. The energy consumed during the trip is only 11\% of the maximum battery energy. We observe some high-frequency jitters in roll which are likely due to numerical issues. 

Fig. \ref{Fig: trajectory} shows the obstacles (red), the convex polygons in the obstacle-free safe region (green), and the trajectory (orange) in cruise. The trajectory deviates from a straight line under the influence of wind and obstacles. The figure also shows a zoomed-in part of the trajectory (marked by a dotted circle) in the vicinity of many obstacles. We observe that the trajectory lies entirely within the safe polygons, avoiding obstacles. Wherever possible, GCS-OCP gives a trajectory passing \textit{in-between} (rather than \textit{around}) obstacles, which helps to reduce the flight time.


\subsection{Comparison of GCS-OCP with Baselines} \label{Subsec: comparison results}
\begin{table*}[htb] 
	\centering 
	\caption{Total flight and computational times for different origins in NYC, altitudes, number of obstacles for GCS-OCP framework and other baselines. The shorter times between D-OCP and GCS-OCP are \textbf{bolded}. `Dist.': distance, `elev.': elevation, `Alt.': cruising altitude, `obs.': obstacles, `Diff.': difference, `DNC': did not converge}
	\label{Tab: results}
	\begin{threeparttable}
		\begin{tabular}{|c|c|c|c|c||c||c|c|c||c|c|c||c|c|} 
			\hline
			\multirow{4}{*}{\textbf{Origin}} & \multirow{4}{*}{\textbf{Dist. to}} & \multirow{4}{*}{\textbf{Origin}} & \multirow{4}{*}{\textbf{\textbf{Alt.,}}} & \multirow{4}{*}{\textbf{\# obs.}} & \multirow{2}{*}{\textbf{BL1}} & \multicolumn{3}{c||}{\multirow{2}{*}{\textbf{D-OCP}}} & \multicolumn{3}{c||}{\multirow{2}{*}{\textbf{GCS-OCP}}} & \multicolumn{2}{c|}{\multirow{2}{*}{\textbf{Diff. in flight time,}}} \\ 
			\multirow{4}{*}{\textbf{\#}} & \multirow{4}{*}{\textbf{JFK, km}} & \multirow{4}{*}{\textbf{elev., m }} & \multirow{4}{*}{\textbf{\textbf{m}}} & & & \multicolumn{3}{c||}{\multirow{2}{*}{}} & \multicolumn{3}{c||}{\multirow{2}{*}{}} & \multicolumn{2}{c|}{\multirow{2}{*}{\textbf{\% of GCS-OCP}}} \\

			& & & & & Flight & Flight & Run- & IPOPT & Flight & Run- & IPOPT & \multirow{2}{*}{BL1} & \multirow{2}{*}{D-OCP} \\  
			& & & & & time, s & time, s & time, s & time, s & time, s & time, s & time, s & & \\ \hline \hline
			
			\multirow{3}{*}{1} &  \multirow{3}{*}{16} & \multirow{3}{*}{30} & 136 & 6 & 480 & 585 & \textbf{434} & \textbf{62} & \textbf{481} & 1194 & \textbf{62} & 0 & 22 \\
			& & & 150 & 1 & 498 & \textbf{569} & 430 & 60 & 578 & \textbf{407} & \textbf{29} & -14 & -2\\
			& & & 163 & 0 & 515 & 610 & \textbf{417} & \textbf{73} & \textbf{609} & 464 & 85 & -15 & 0\\ \hline		
			\multirow{5}{*}{2} & \multirow{5}{*}{22} & \multirow{5}{*}{71} & 136 & 112 & 565 & DNC & DNC & DNC & \textbf{630} & \textbf{625} & \textbf{17} & -10 & DNC\\
			& & & 165 & 48 & 603 & DNC & DNC & DNC & \textbf{710} & \textbf{645} & \textbf{20} & -15 & DNC\\
			& & & 194 & 26 & 641 & DNC & DNC & DNC & \textbf{756} & \textbf{454} & \textbf{16} & -15 & DNC\\ 
			& & & 251 & 7 & 716 & 861 & \textbf{407} & 62 & \textbf{830} & 663 & \textbf{16} & -14 & 4\\ \hline		
			\multirow{5}{*}{3} & \multirow{5}{*}{23} & \multirow{5}{*}{76} & 136 & 253 & 587 & DNC & DNC & DNC & \textbf{731} & \textbf{1587} & \textbf{25} & -20 & DNC\\
			& & & 152 & 168 & 608 & DNC & DNC & DNC & \textbf{717} & \textbf{1207} & \textbf{17} & -15 & DNC\\
			& & & 168 & 33 & 629 & 768 & \textbf{568} & 85 & \textbf{730} & 791 & \textbf{14} & -14 & 5\\ 
			& & & 200 & 6 & 671 & 850 & \textbf{387} & 35 & \textbf{821} & 430 & \textbf{15} & -18 & 4\\ \hline		
			\multirow{4}{*}{4} & \multirow{4}{*}{24} & \multirow{4}{*}{82} & 136 & 58 & 591 & 720 & \textbf{936} & 179 & \textbf{644} & 2430 & \textbf{115} & -8 & 12\\
			& & & 157 & 23 & 619 & 757 & \textbf{875} & 164 & \textbf{677} & 2229 & \textbf{131} & -9 & 12\\
			& & & 178 & 9 & 646 & 752 & \textbf{466} & 90 & \textbf{709} & 1890 & \textbf{44} & -9 & 6\\ 
			& & & 220 & 0 & 701 & 865 & \textbf{383} & \textbf{52} & \textbf{864} & 432 & 57 & -19 & 0\\ \hline	
			
			\multirow{3}{*}{5} & \multirow{3}{*}{24} & \multirow{3}{*}{35} & 136 & 13 & 628 & 738 & \textbf{475} & \textbf{64} & \textbf{677} & 1465 & 67 & -7 & 9\\
			& & & 149 & 5 & 645 & 830 & \textbf{414} & \textbf{33} & \textbf{707} & 1224 & 74 & -9 & 17\\
			& & & 174 & 0 & 677 & \textbf{842} & \textbf{409} & 62 & \textbf{842} & 432 & \textbf{55} & -20 & 0\\ \hline
			
				
		\end{tabular}
	\end{threeparttable}
\end{table*}

We compare our GCS-OCP framework with two baselines: i) a `back-of-the-envelope' method (\textbf{BL1}) which ignores wind and obstacles and assumes phase-wise constant eVTOL airspeeds (1.5 ms$^{-1}$ in ascent and descent, and 50.4 ms$^{-1}$ in cruise \cite{Chati2025}) to give straight line trajectories; and ii) \textbf{D-OCP} described in Sec. \ref{Sec: framework}.
Table \ref{Tab: results} shows the total end-to-end flight times for the different methods for various trip scenarios. First, we see that for the same origin, the total flight time (as given by GCS-OCP) increases with altitude. Thus, despite more obstacles, cruising at a lower altitude results in a lower total flight time. Second, we observe that ignoring realistic constraints like wind, obstacles, or accurate flight dynamics (BL1) underestimates the total flight time, by as much as 20\%, compared to that from GCS-OCP. Third, D-OCP may give trajectories with higher total flight times, by as much as 22\%, than GCS-OCP.  Lastly, we observe that unlike GCS-OCP, D-OCP may simply fail to converge in the presence of a large number of obstacles (for example, 250) due to the nonconvex nature of its obstacle avoidance constraints. Thus, GCS-OCP is more scalable than D-OCP.

\subsection{Computational Performance} \label{Subsec: computational performance}
\begin{figure}[b]
	\centering
	\includegraphics[width=\columnwidth, trim=0cm 0cm 0cm 0cm, 
	clip]{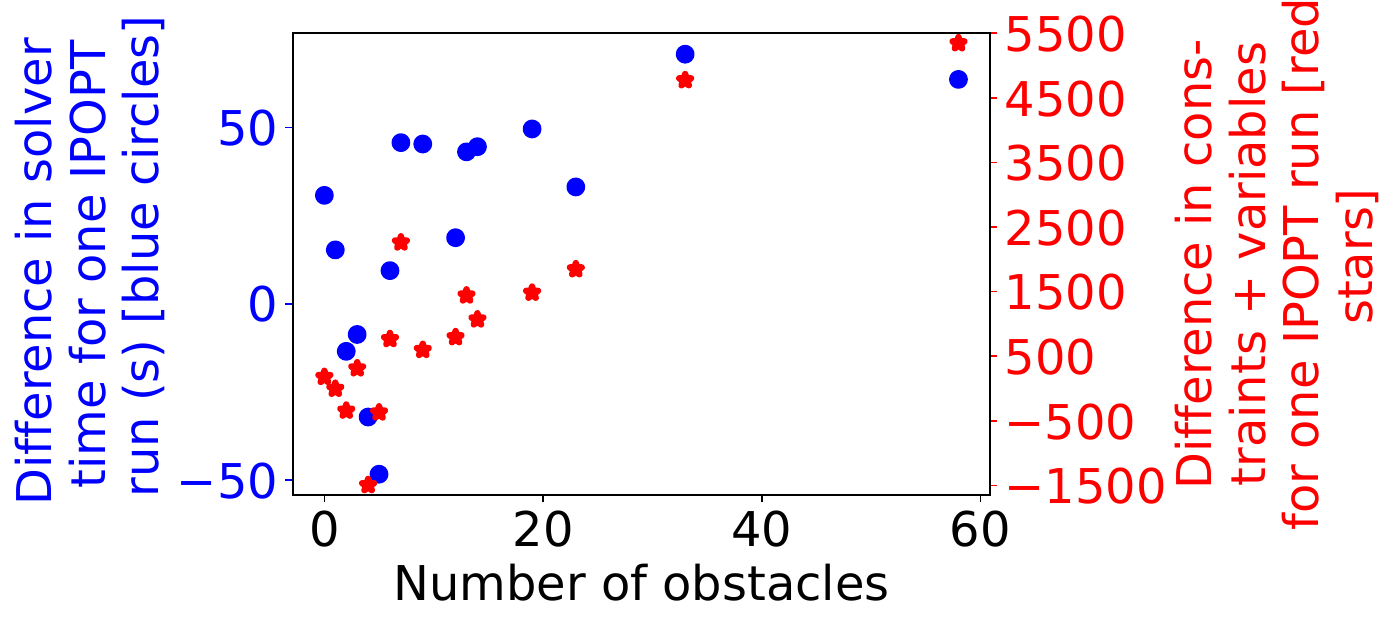}
	\caption{\label{Fig: solver performance} Comparison of computational performance between D-OCP and GCS-OCP for one IPOPT solver run.}
\end{figure}

Table \ref{Tab: results} also shows the total runtimes of D-OCP and GCS-OCP. For D-OCP, the runtime corresponds to only running the OCP over multiple cruise intervals $N_\textrm{cr}$ till convergence. However, for GCS-OCP the total runtime corresponds to the times required to i) get the convex polygons using the IRIS algorithm, ii) get candidate paths using the GCS-based planner, and iii) run the OCP solver over multiple candidate paths and cruise intervals. Thus, GCS-OCP is overall computationally slower than D-OCP (whenever D-OCP converges).
However, it is also instructive to compare the computational performance for only one run of the IPOPT solver used for OCP. Fig. \ref{Fig: solver performance} shows the difference of the solver times, and the number of NLP variables and constraints, between D-OCP and GCS-OCP for different number of obstacles. 
We observe that at low number of obstacles, D-OCP has less constraints and variables than GCS-OCP and can be faster. However, as the number of obstacles increases, the number of variables and constraints and the solver time for D-OCP exceed those for GCS-OCP. Thus, enforcing the trajectory to \textit{pass} through convex safe regions rather than \textit{avoid} obstacles is more computationally efficient for large number of obstacles. 

\subsection{Limitations} \label{Subsec: limitations}
We discuss some limitations of our work. First, the long solution times make GCS-OCP suitable for offline trajectory planning only (and not for real-time trajectory re-planning). Second, like most NLP problems, the solution from GCS-OCP is not guaranteed to be truly optimal. Third, we have only considered a static, uniform wind field and not a more realistic dynamically varying, nonuniform wind field with a vertical component. Fourth, we have considered only stationary obstacles and not moving obstacles (for example, other eVTOLs in flight). 
%

\section{Conclusions} \label{Sec: conclusions}
We develop a hybrid GCS-OCP framework to design the end-to-end eVTOL trajectory for UAM operations in the presence of realistic constraints imposed by complex flight dynamics, wind, building obstacles, and operational bounds. We show that not considering these constraints can underestimate the actual flight time by as much as 20\%, which underscores the usefulness of our framework to plan for UAM operations accurately. We also show that, by enforcing the trajectory to pass through convex flight-safe regions, our framework gives a shorter (and often more computationally efficient) trajectory than a more direct, but nonconvex, OCP formulation that enforces the trajectory to avoid obstacles.
Future directions to our work could include: i) developing a real-time framework, ii) considering dynamic wind fields, iii) including moving obstacles, and iv) using other objectives (like energy consumed) for the trajectory design. 

\bibliographystyle{IEEEtran}
\bibliography{bibfile} 

\end{document}